\documentclass[twocolumn]{revtex4}

\usepackage {graphicx}

\begin{document}

\title{Coherent X-ray Diffractive Imaging; applications and limitations}

\author{S.~Marchesini}
\affiliation{Lawrence Livermore National Laboratory, 7000 East Ave., Livermore CA 94550 USA.}

\author{H.~N.~Chapman}
\affiliation{Lawrence Livermore National Laboratory, 7000 East Ave., Livermore CA 94550 USA.}
\author{S.~P.~Hau-Riege}
\affiliation{Lawrence Livermore National Laboratory, 7000 East Ave., Livermore CA 94550 USA.}
\author{R.~A.~London}
\affiliation{Lawrence Livermore National Laboratory, 7000 East Ave., Livermore CA 94550 USA.}
\author{A.~Sz\"oke}
\affiliation{Lawrence Livermore National Laboratory, 7000 East Ave., Livermore CA 94550 USA.}
\author{H.~He}
\affiliation{Advanced Light Source, Lawrence Berkeley Lab, 1 Cyclotron Rd., Berkeley, CA 94720. USA.}
\author{M.~R.~Howells}
\affiliation{Advanced Light Source, Lawrence Berkeley Lab, 1 Cyclotron Rd., Berkeley, CA 94720. USA.}
\author{H.~Padmore}
\affiliation{Advanced Light Source, Lawrence Berkeley Lab, 1 Cyclotron Rd., Berkeley, CA 94720. USA.}
\author{R.~Rosen}
\affiliation{Advanced Light Source, Lawrence Berkeley Lab, 1 Cyclotron Rd., Berkeley, CA 94720. USA.}
\author{J.~C.~H.~Spence}
\affiliation{Department of Physics and Astronomy, Arizona State University, Tempe AZ 85287-1504, USA.}
\author{U.~Weierstall}
\affiliation{Department of Physics and Astronomy, Arizona State University, Tempe AZ 85287-1504, USA.}

\begin{abstract} 
The inversion of a diffraction pattern offers 
aberration-free diffraction-limited 3D images without the resolution and 
depth-of-field limitations of lens-based tomographic systems, the only 
limitation being radiation damage. We review our experimental results, 
discuss the fundamental limits of this technique and future plans.
\end{abstract}

\preprint{UCRL-JC-155105}
\maketitle

\section{Introduction}

Three ideas, developed over half a century, have now converged to provide a 
working solution to the non-crystallographic phase problem. In this paper we 
outline some applications and our development of this solution, which 
provides a method for lensless, diffraction-limited, aberration-free X-ray 
imaging of nano-objects in three-dimensions at high resolution. We suggest 
the acronym CXDI (Coherent X-ray Diffractive Imaging) for this field. We 
also discuss the fundamental limitations on resolution set by radiation 
damage, and some new approaches to this problem based on femtosecond 
diffraction. 

The ideas start with Sayre's 1952 observation that Bragg diffraction 
undersamples diffracted intensity relative to Shannon's theorem \cite{sayre:1952}. 
Secondly, the development of iterative algorithms with feedback in the early 
nineteen-eighties produced a remarkably successful optimization method 
capable of extracting phase information from adequately sampled intensity 
data \cite{fienup:1982}. Finally, the important theoretical insight that these 
iterations may be viewed as Bregman Projections in Hilbert space has allowed 
theoreticians to analyze and improve on the basic Fienup algorithm 
\cite{elser:2003}. A parallel development was the real space algorithm for ``phase 
recovery'' in crystallography \cite{szoke:1997} and non-periodic diffraction 
\cite{hauriege:2003}. 

Experimental work started for X-rays with the images of lithographed 
structures reconstructed from their soft-X-ray diffraction patterns by Miao 
et al. in 1999 \cite{miao:1999}. More recently we have seen higher resolution X-ray 
imaging \cite{miao:2002}, 3D imaging \cite{miao:2002, williams:2003}, the 
introduction of algorithms which do not require a conventional 
lower-resolution image to provide the boundary (support) of the object 
\cite{he:acta:2003, he:prb:2003, marchesini:2003} and the striking 
atomic-resolution image of a single nanotube reconstructed from an 
experimental electron micro-diffraction pattern (an example of the more 
general CDI method) \cite{zuo:2003}. 

The rapid growth of nanoscience (described recently as ``the next industrial 
revolution") has produced an urgent need for techniques capable of revealing 
the internal structure, in three dimensions, of inorganic nanostructures and 
large molecules which cannot be crystallized (such as the membrane proteins 
of vital importance for drug delivery). Scanning probe methods are limited 
to surface structures, and the electron microscope can provide atomic 
resolution images of projections of crystalline materials in thicknesses up 
to about 50nm, or tomography of macromolecular assemblies and inorganics at 
lower resolution. No technique at present can provide three-dimensional 
imaging at nanometer resolution of the interior of particles in the micron 
size range. The development of such a method would have a decisive impact on 
several fields of science, and would be a timely enabling technology for the 
Molecular Foundary at Lawrence Berkeley National Laboratory, the Linac 
Coherent Light Source (LCLS) at Stanford and other nanoscience user facility 
initiatives. Briefly, we foresee initial applications for CXDI (in the 
0.5--4kV X-ray range) in materials science as follows: 1. The visualization 
of the internal labyrinth structure of the new mesoporous framework 
structures (eg. glassy foams, now finding uses for molecular sieves and 
hydrogen storage; 2. Imaging the complex tangles of dislocation lines which 
are responsible for work-hardening; 3. Imaging the cavities within duplex 
steels, responsible for their very high uniform extension; 4. 3D imaging 
defect structures in magnetic multilayers; 5. The tomographic imaging of 
misfit dislocations at interfaces, free of the thin-film elastic relaxation 
processes which distort the images obtained by transmission electron 
microscopy; 6. Imaging of the three-dimensional arrangement of Orowan 
dislocation loops which, by entanglement with particles, provide the 
dispersion-hardening of copper alloys; 7. The imaging of precipitates in 
metal-matrix composite materials; 8. The imaging of electronic device 
elements for future computing schemes, such as quantum computing. This 
application is particularly promising, since the ability to prepare the 
elements lithographically provides the a-priori information needed to solve 
the phase problem. In life sciences such a technique is needed to determine 
the bulk (internal) structure of assemblies of macromolecules (molecular 
machines), protein complexes, and virus particles at a resolution sufficient 
to recognize known proteins and determine their relationships to each other. 
The apparatus we have developed for CXDI is reviewed in section 3 and 
reconstruction methods are described in section 4. Section 5 summarizes data 
on the radiation damage limitations to resolution, and the theoretical basis 
for these limits, while Section 6 describes a method for overcoming those 
limits by using ultrafast x-ray pulses. We begin by describing a new 
algorithm for CXDI which avoids the need for \textit{a priori} knowledge of the object 
support.

\section{The CXDI technique}

A CXDI experiment consists of three steps: (a) the sample is illuminated by 
monochromatic coherent x-rays and a recording is made of a single 
diffraction pattern (for 2D) or a tilt series (for 3D); (b) the phases of 
the pattern are recovered from the measured intensities using established 
phase-retrieval algorithms; (c) the unknown object is then recovered by 
Fourier inversion. In the Gerchberg-Saxton-Fienup scheme one starts with 
random phases, which lead to noise when transformed from reciprocal to real 
space. One then imposes the ``finite support'' constraint (namely that there 
must be a blank frame around the specimen), before transforming back to 
reciprocal space. In reciprocal space the phases so generated are combined 
with the measured diffraction magnitudes to start the next iteration. After 
a large number of iterations, in most cases the object emerges from the 
noise. Our novel improvement is that the estimate for the object support is 
continually updated by thresholding the intensity of the current object 
reconstruction \cite{marchesini:2003}. We start from a threshold of the 
transform of the diffraction pattern and as the iterations progress the 
support converges to a tight boundary around the object. This, in turn, 
improves the image reconstruction, which gives a better estimate of the 
support. An example of the reconstruction of a simulated diffraction pattern 
produced by a cluster of gold balls is shown in the movie (Fig. \ref{fig3}) together 
with the support.

The algorithm does not require any ``atomicity" constraint provided by the 
gold balls as demonstrated by the reconstruction of a greyscale image. The 
algorithm also successfully reconstructs complex objects (those that cause 
large variations in the phase of the exit wavefield in two dimensions), 
which hitherto have been experimentally difficult to reconstruct. This opens 
up the possibility of image reconstruction from microdiffraction patterns, 
where the illumination is tightly focused on the object.

\begin{figure*}[htbp]
\centerline{
\includegraphics[height=2in]{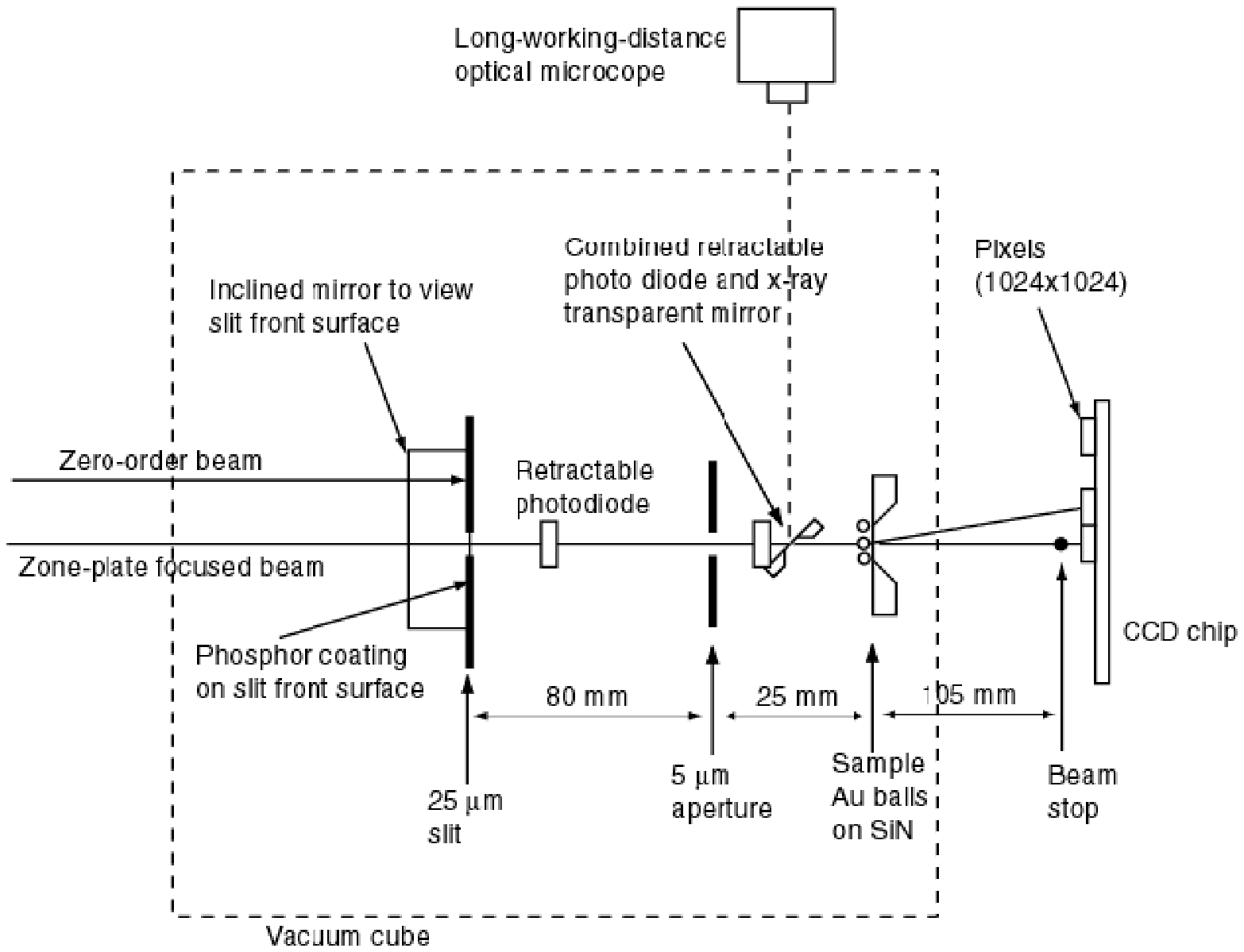}
\includegraphics[height=2in]{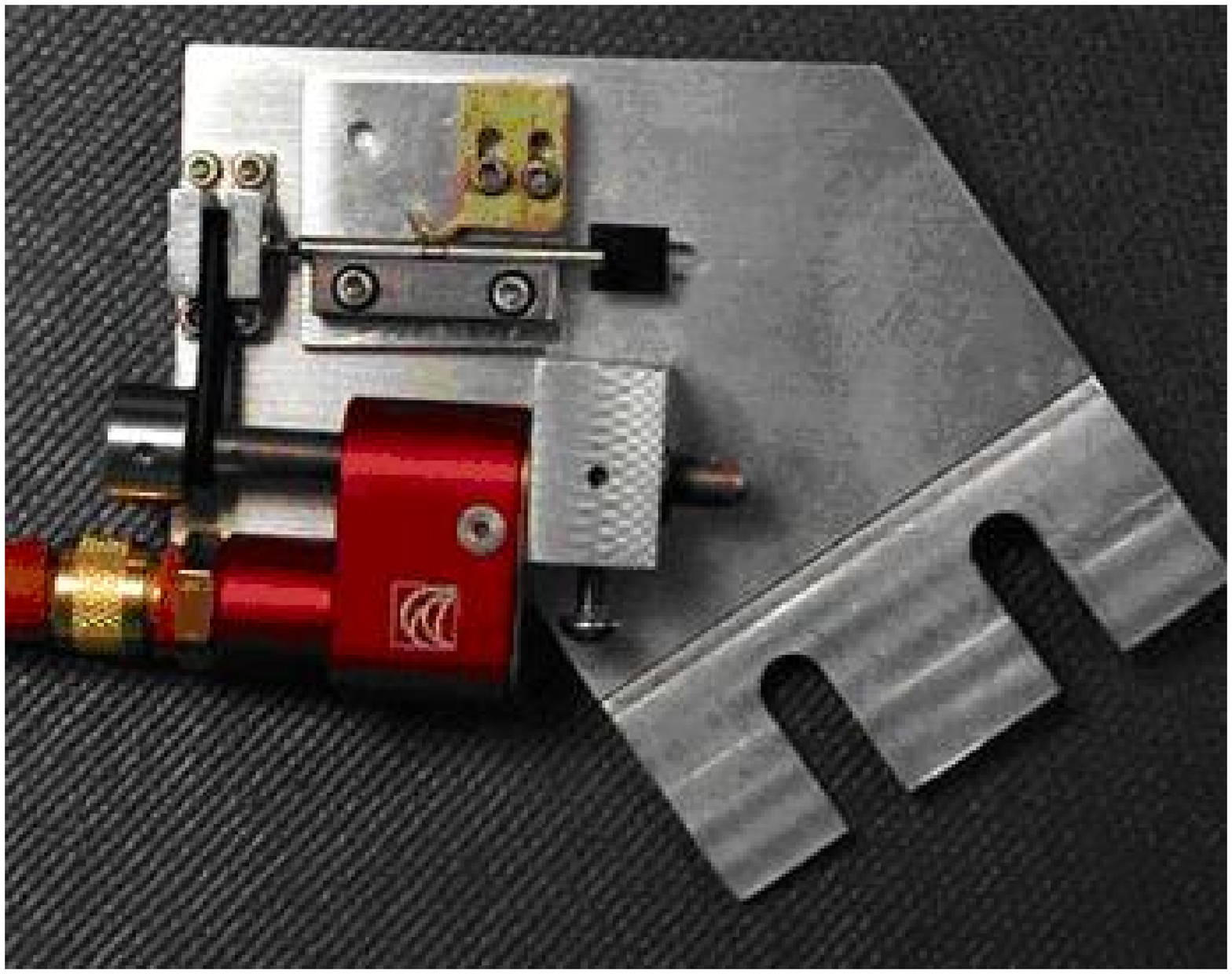}
}
\caption{Experimental chamber layout (left) and rotation device for tomographic 
recordings (right).}
\label{fig1}
\end{figure*}

\begin{figure*}[htbp]
\includegraphics[height=1.5in]{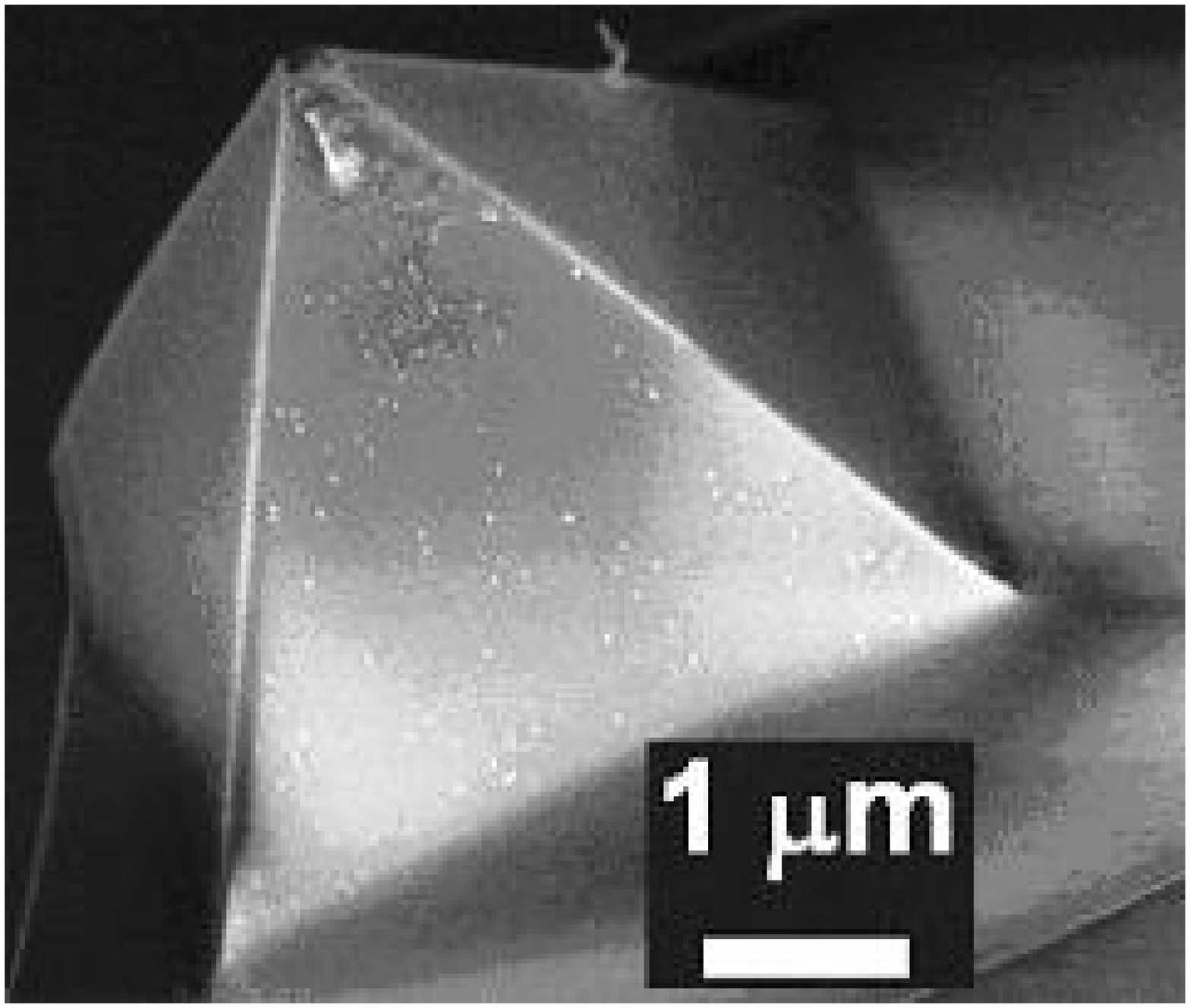}
\hspace{1in}
\includegraphics[height=1.5in]{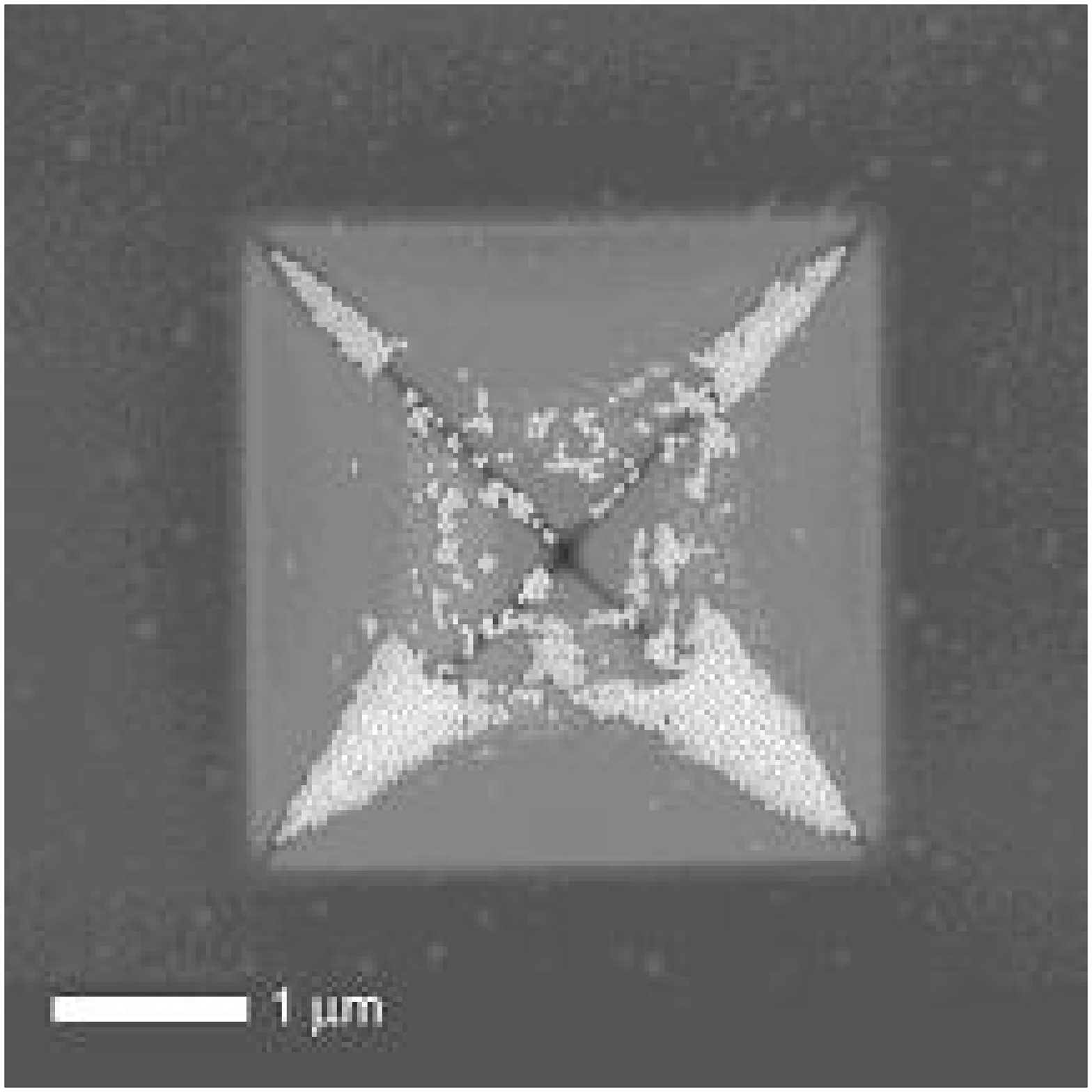}
\caption{
Three-dimensional test objects fabricated for x-ray diffraction imaging.  The left picture shows a SEM perspective image of a silicon nitride pyramid membrane, and the right picture shows a top view of a similar pyramid that has been decorated with 50 nm diameter gold spheres.  The silicon nitride is 100 nm thick, and the pyramid is hollow.  These objects are ideal for testing since they can be well characterized in the SEM, have extent in all three dimensions, and can be treated as an analogue to a molecule}
\label{fig2}
\end{figure*}

\section{CXDI experiments at the ALS}

Our experiments in coherent diffraction began with trials using electron 
\cite{weiestrall:2002} and visible-light \cite{spence:2002} optics and continued with 
CXDI at the Advanced Light Source (ALS). The experiments used the ``pink" 
beam at beam-line 9.0.1 \cite{howells:2002} which is fed by a 10-cm-period 
undulator operating in third harmonic with deflection parameter ($K$) equal to 
1.2 and delivering 588 eV (2.11 nm) photons. Features of the beam line 
include a 0.5 $\mu $m-thick, 750 $\mu $m-square Be window to separate the 
UHV beam line from the low-vacuum sample environment, a monochromator 
consisting of an off-axis segment of a zone plate and the diffraction 
experiment itself (Fig. \ref{fig1}). 

The x-ray coherence length $l_{c} $ must be greater than the maximum path 
difference between any pair of interfering rays, i. e. $l_{c} >w\theta 
_{\max } $, where $w$ is the width of the sample and $\theta _{\max } $ is the 
maximum diffraction angle. For our geometry and wavelength, $\theta _{\max } 
$= 0.12 radian and the resolution limit is 8.4 nm. For the 5~$\mu $m 
aperture (effectively the monochromator exit slit) shown in Fig. \ref{fig1}, the 
resolving power is about 500, the coherence length is then 1$\mu $m and the 
maximum sample illumination area 8$\times $8 $\mu $m$^{2}$. Similarly the 
(spatial) coherence patch provided by the 5 $\mu $m aperture is 10$\times 
$10 $\mu $m$^{2}$. Allowing for an empty (but still coherently illuminated) 
band around the sample, its allowed size is thus $<$ 4$\times $4 $\mu 
$m$^{2}$.

We consider now the sampling of the diffraction pattern. The Shannon 
interval for frequency-space sampling of the \textit{intensity} is
 $1 \mathord{\left/ 
{\vphantom {1 {\left( {2w} \right)}}} \right. \kern-\nulldelimiterspace} 
{\left( {2w} \right)}=\Delta \mathord{\left/ {\vphantom {\Delta {\lambda 
z}}} \right. \kern-\nulldelimiterspace} {\lambda z}$ where $z$ is the 
sample-to-detector distance and \textit{$\Delta $} is the detector-plane 
increment (a 25 $\mu 
$m CCD pixel in our case). For our \textit{$\lambda $} and $z$ values this 
also leads to a maximum 
sample width of 4 $\mu $m. This is correct (Shannon) sampling of the 
diffraction-plane \textit{intensity} and twofold \textit{over}sampling 
in each direction of the 
diffraction-plane wave amplitude. Note that these limits on the sample size 
arising from coherence and sampling considerations are not the only ones in 
effect. 

We have carried out three series of experiments, all using test samples made 
from 50 nm gold balls. The first \cite{he:acta:2003} demonstrated the basic 2D 
technique with image reconstruction using a support function determined by 
scanning electron microscopy. The second \cite{he:acta:2003} used a sample 
intentionally prepared in two separated parts, and reconstruction was 
achieved using information from the 2D diffraction pattern alone. The third 
series used a miniature sample-rotation device to collect several 
tomographic data sets. The picture of the device shown in Fig. \ref{fig1} (picture 
width = 7 cm) shows the sample and its rotation spindle and driver. The 
small black square is the frame of a Si$_{3}$N$_{4}$ window on which the 3D 
sample is deposited in a 2.5-$\mu $m-wide microfabricated tetrahedron (Fig. \ref{fig2}). 
Not shown is an angular Vernier scale that was used to measure the 
rotation angle. Using this apparatus, a set of 150 views with at least a 100 
second exposure time per view required about 10 hours. The 3D data generated 
by the object shown in Fig. \ref{fig2} are still being analysed.

\section{Reconstructions }

Samples were made by placing a droplet of solution containing `gold 
conjugate' colloidal gold balls on a silicon nitride window (thickness 
100~nm) and allowing it to dry. The gold balls formed several single 
layered (2D) clusters on the SiN membranes, as imaged by a field-emission 
scanning electron microscope (SEM).

In the first experiment the use of a Si$_{3}$N$_{4}$ window of the order of 
5 microns width ensured that the sample was isolated and of the required 
size \cite{he:acta:2003}. The structure contained at least one isolated ball 
generating a `reference wave' which interfered with the signals from other 
clusters to form a hologram.

\begin{figure}[htbp]
\includegraphics[height=2in]{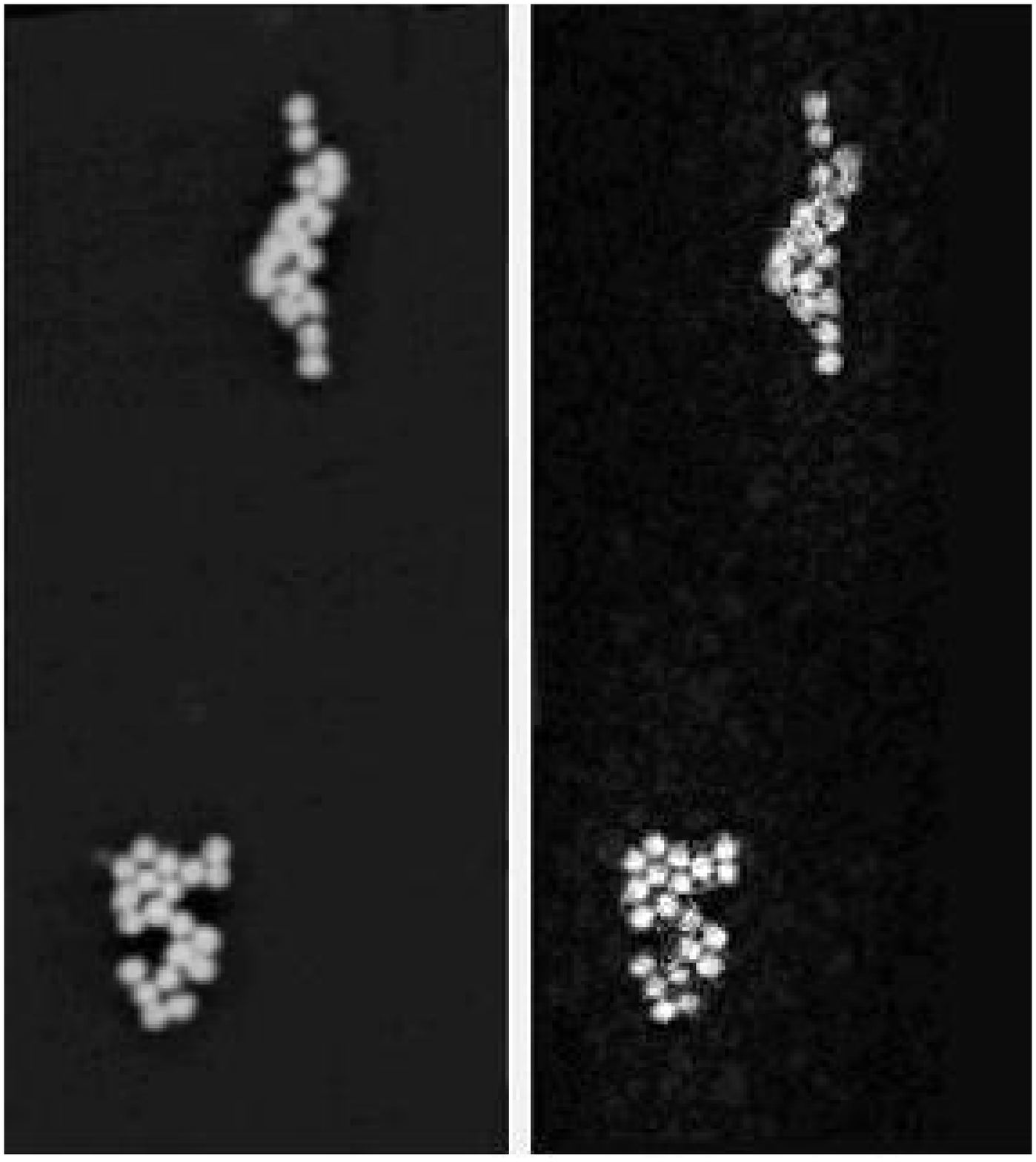}
\includegraphics[height=2in]{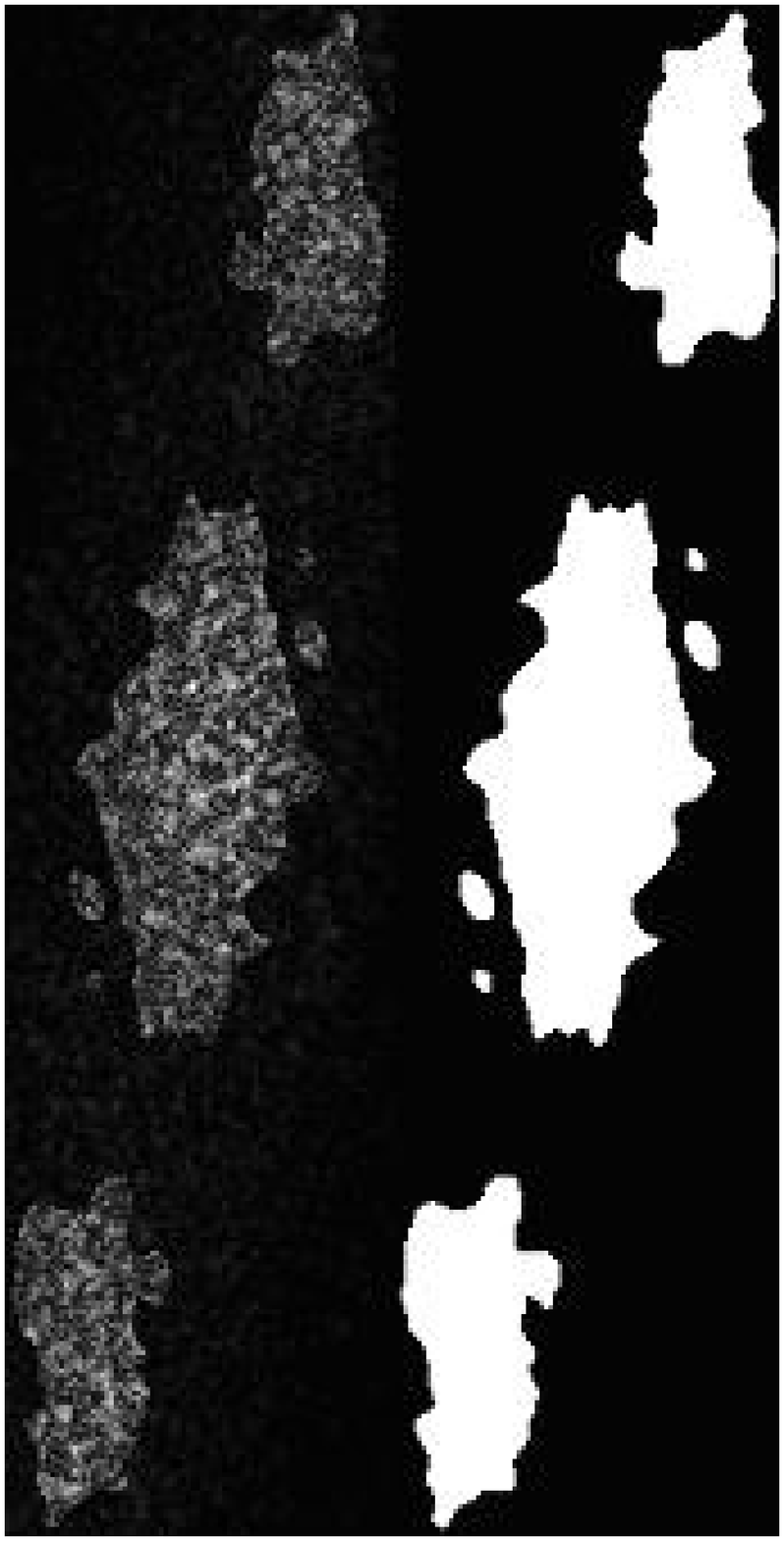}
\caption{
Comparison of reconstructed soft X-ray image (middle) and SEM images of gold ball clusters (left). 
Each ball has a diameter of 50nm 
[H. He, et al. Phys. Rev. B, 67, 174114 (2003)].  Also shown (right) is a movie (1.1 MB) of the reconstruction as it
 iterates.  Each frame of the movie displays the current estimate of the image intensity on the left and the image 
support on the right \cite{marchesini:2003}.}
\label{fig3}
\end{figure}

The autocorrelation function obtained by Fourier transforming the intensity 
of the diffraction pattern included an image of every cluster convolved with 
the single ball, and these images formed a faithful representation of the 
structure in real space, without iterative processing. Not all the clusters 
could be imaged this way, since some of the intra-cluster distances were 
overlapping. The complex transmission function of the Si wedge at the corner 
of the window generated a complex object that was difficult to reconstruct 
by phase retrieval. For real objects, we have seen in simulations that using 
a support of triangular shape was sufficient to obtain the image. In the experimental 
case, we had to use a support obtained from a low resolution image of the 
object obtained by scanning electron microscopy (SEM). In the next 
experiment the object was isolated by means of sweeping the sample particles 
with an Atomic Force Microscope (AFM). The reconstruction has been obtained 
using the standard HIO algorithm \cite{fienup:1982} with missing data due to the beam 
stop and its supports left unconstrained during the iterations. Rather than 
rely on a low-resolution secondary imaging method to obtain an estimate of 
the support, we obtained an estimate of the object support from the support 
of its autocorrelation. The object support was then adjusted manually as the 
reconstructed image became clearer \cite{he:prb:2003}. A recently developed 
algorithm allows both initial support selection and adjustment to be 
performed automatically \cite{marchesini:2003}, extending the technique to 
objects other than a few separated clusters. This algorithm has been 
successfully applied on simulated 3D diffraction patterns, and it is being 
tested on the experimental 3D data recently recorded. We are also 
investigating the use of the real space algorithm \cite{hauriege:2003} to reconstruct 
the clusters in Fig. \ref{fig3} and the 3D data. While this algorithm does require 
prior information in the form of a low-resolution image, it may be 
advantageous for the reconstruction from sparse, irregular, incomplete and 
noisy data.

\section{Dose and Flux limitations to performance}

In order to assess the promise of CXDI we have carried out calculations 
intended to determine the fluence (total photons per unit area) and dose 
(absorbed energy per unit mass) required to make a 3D CXDI image at given 
resolution. The basis of the calculation was the so-called ``dose 
fractionation'' theorem \cite{hegerl:1976, mcewen:1995}. 
The theorem states that the 
radiation dose required to measure the strength of a voxel is the same for a 
multi-view 3-D experiment (with appropriate reconstruction algorithms) as it 
is for measurement of the voxel alone, provided that the voxel size and the 
statistical accuracy are the same in both cases. The results are therefore 
based on a calculation of diffraction by a single voxel. The conclusions 
were as follows:

\begin{figure*}[htbp]
\centerline{
\includegraphics[width=3.5in]{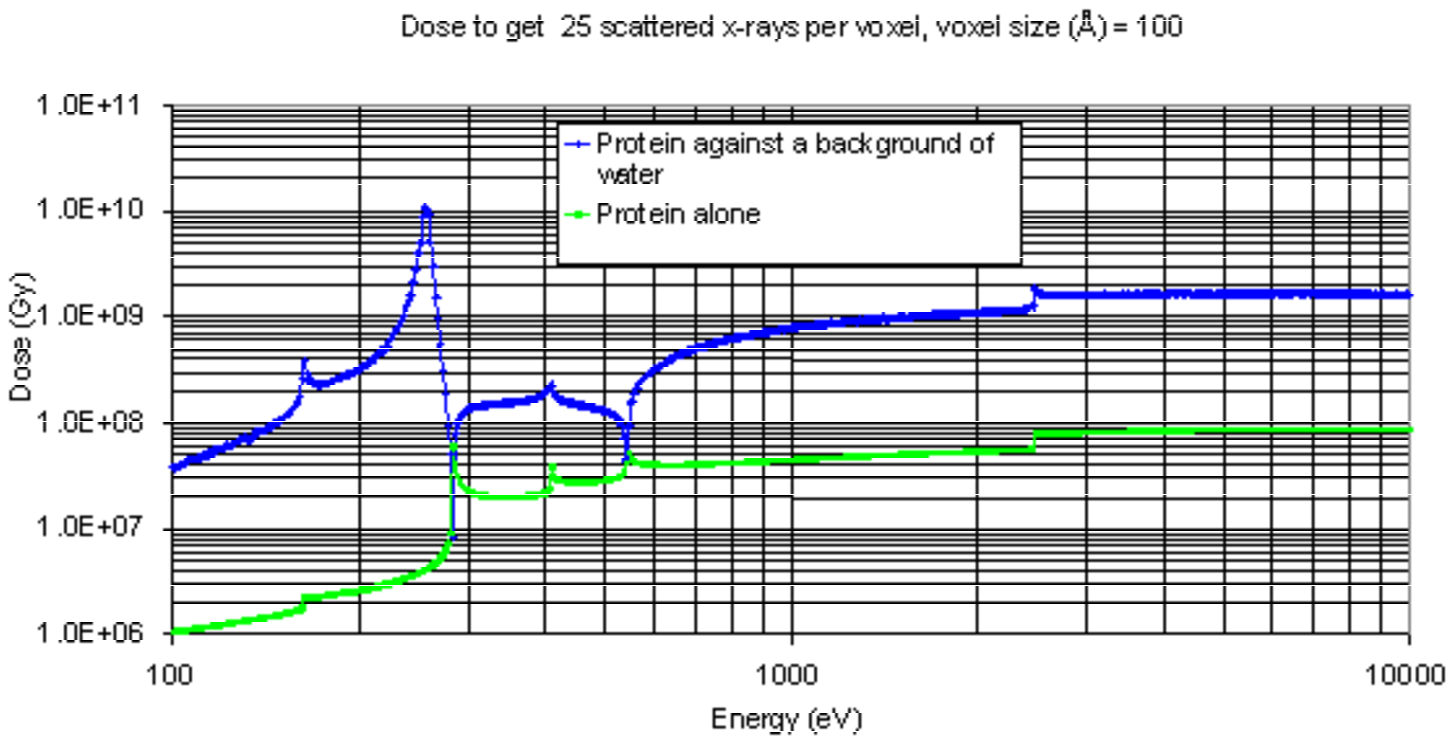}
\includegraphics[width=3.5in]{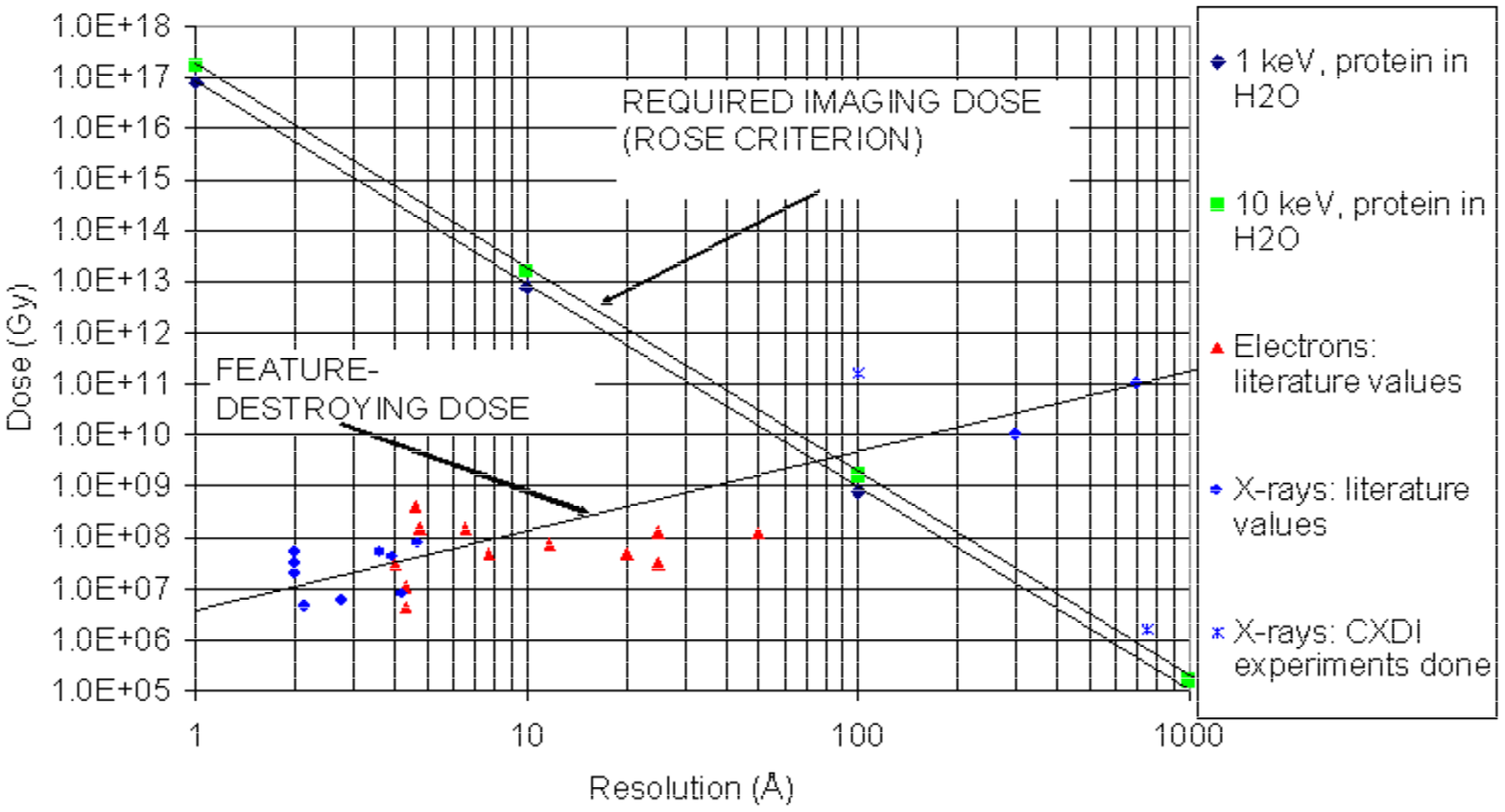}
}
\caption{(left) Plot of dose against x-ray energy.  (right) Plot of dose against resolution.}
\label{fig4}
\end{figure*}

\begin{enumerate}
\item The dose ($D$) and fluence ($N$) needed to produce $P$ scattered 
x-rays per voxel into a detector collecting the angle required for 
resolution $d$, are given by 
$N=P/ \sigma _{s}$ and  $D=\mu h\nu P/ \varepsilon \sigma_s$ 
where $\mu $ is the absorption coefficient, $h\nu $ the 
photon energy and $\varepsilon$ the density. 
\item The scattering cross section $\sigma _{s} $ of the voxel of size 
$d\times d\times d$ is given by 
$\sigma _{s} =r_{e}^{2} \,\lambda ^2\,\left| \rho \right|^2\,d^4$ 
(see also \cite{henke:1955}) where $r_{e} $ is the classical 
electron radius, $\lambda $ the wavelength and $\rho$ 
the complex electron density. 
\item For resolution $d$, the dose and fluence scale as $d^{-4}$
\item For light elements \textit{$\rho $} is fairly constant above about 
1 keV so the energy 
dependence of the dose is expected to be quite flat in the range 1- 10 keV.
\item $N$ is dominated by the cross section and scales with the square 
of photon energy. Moreover, the coherent power of a source of brightness $B$ 
is $B\left(\lambda/2 \right )^2$. This implies a fourth-power 
fluence penalty for increasing the x-ray energy.
\item Therefore one should use the lowest possible x-ray energy 
consistent with (roughly) $\lambda <d/2$
\item The dose for detecting a 10 nm protein feature against a background of water according to the Rose criterion \cite{rose:1948} is shown in Fig. \ref{fig4}. The required 
imaging dose in the energy range 1-10 keV is roughly $10^9$ Gy
\end{enumerate}

\textit{Quantitative dose limit:} 
to obtain an estimate of the resolution limit we have plotted a variety of 
literature values of the dose needed to destroy biological features as a 
function of feature size (Fig. \ref{fig1}). Also plotted is the required imaging 
dose. The tentative conclusion from the graph is that the resolution of the 
crossover, i.e. about 10 nm, should be possible for unlabelled life-science 
samples although for material-sciences samples the radiation tolerance (and 
thus the resolution) can be much higher.

\textit{Quantitative fluence limit:} in our latest experiment at ALS, we collected a full 3D data set at a 
resolution that we believe to be around 10 nm (although this is not yet 
supported any reconstructions) in about 10 hours. We project that a beam 
line \textit{optimized} for this experiment operating on the ALS after its planned performance 
upgrade would collect diffraction data about $10^4$ times 
faster than now. From 2, above, this should allow the step from 10 nm to 
around 1 nm resolution for sufficiently radiation-resistant samples.

\section{Femtosecond CXDI, beyond the radiation-damage limit}

A way of overcoming the radiation damage limit in x-ray imaging is to use 
pulses of x-rays that are shorter in duration than the damage process 
itself. This idea of flash imaging, first suggested by Solem \cite{solem:1982, 
solem:1984}, has been proposed to be extended all the way to atomic resolution using 
femtosecond pulses from an x-ray free-electron laser (XFEL) \cite{neutze:2000}. The 
methodology of CXDI could be used in this case to image single molecules 
\cite{neutze:2000, szoke:1999, miao:2001}. 
The general concept for imaging non-periodic 
samples is to inject reproducible samples (macromolecules, complexes, or 
virus particles) into the XFEL beam, to intersect one particle (and record 
one diffraction pattern) per pulse. 

In the general XFEL experiment the particle orientation in three dimensions 
will be random and unknown, and the individual diffraction patterns will be 
noisy (especially at the highest diffracted angles where the 
highest-resolution orientation information resides). General ideas and 
methods, developed for ``single-particle'' reconstructions in cryo-electron 
microscopy \cite{franck:1996, vanheel:2000}, for extracting tomographic information 
from a huge ensemble of randomly oriented noisy images can be applied here. 
Just as in single-particle EM, the limitations to image resolution are the 
ability to sort patterns into classes of like-oriented particles, so they 
can be averaged to improve signal to noise, and the reproducibility of 
particles. A statistical analysis \cite{huldt:2003} has shown that 
signal levels much 
less than a single count per pixel are adequate to be able to classify 
diffraction patterns. Even so, due to the exceedingly small scattering cross 
section of a single macromolecule, pulse fluences of $>$10$^{12}$ photons 
per (100 nm)$^{2}$ are required to achieve the required signal levels. This 
corresponds to a dose per particle of $>10^{14}$ Gray. The dose could be 
reduced substantially if methods could be employed to orient the particle in 
free space (even if only along a single axis) or if symmetric nanoclusters 
of particles could be formed.

We have performed calculations to determine, given the required fluence, the 
longest pulse duration possible to acquire the diffraction pattern before 
disruption of the electron density of the particle at the atomic scale. The 
calculation uses a hydrodynamic method treating the electrons and the ions 
as two separate fluids that interact through the Coulomb force and 
ionization processes, the assumption being that a large enough macromolecule 
can be treated as a continuum. Although it does not treat the atomic motions 
as accurately as the molecular dynamics model \cite{neutze:2000}, 
the hydrodynamic 
model is computationally faster and it also enables the inclusion of 
additional physics effects, in particular the inclusion of the 
redistribution of free electrons throughout the molecule.

The dominant interactions between the matter and x-ray pulse are $K$-shell 
photoionization, (producing damage) and elastic scattering producing the 
diffraction pattern. Following x-ray absorption in about 2 to 10~fs 
\cite{mcguire:1969}, the $K$-shell holes decay by emitting Auger electrons with energies 
in the range of 250 to 500~eV \cite{furguson:1989}. As the photo and Auger electrons 
escape they leave behind a molecule with increasing charge and the growing 
population of low energy secondary electrons becomes trapped and grows still 
further by collisional ionization (causing the bulk of the damage).

The trapped electrons quickly relax in energy and position to form a 
neutralizing cloud around the positively charged ions. The particle assumes 
roughly a two-zone structure, consisting of a neutral core and a positively 
charged outer shell similar to Debye shielding. On a longer timescale of 
several 10's fs, a macroscopic motion of the whole molecule, called a 
Coulomb explosion, takes place, leading to an outward radial motion of the 
ions. The ion motion is greatest within the charged outer layer of the 
molecule. In the core of the molecule the electron temperature is highest, 
with the greatest amount of ionization and blurring of the electron density. 
Preliminary results from the hydrodynamic model indicate that collisional 
ionization is very rapid and may limit the maximum pulse length to a value 
smaller than the 10-20 fs inferred by considering only photoionization, 
Auger ionization and atomic motion. The limits imposed by collisional 
ionization might be overcome by developing a method to reconstruct atomic 
positions from partially ionized atoms. 

To date it has not been possible to acquire experimental data on the rate of 
the Coulomb explosion and the effect on electron density of ultrafast 
high-intensity x-ray pulses. We plan to investigate this at the Tesla Test 
Facility Phase 2 \cite{tesla:2003}, a soft-x-ray FEL that will be operational in 2004 
initially at a wavelength of 20 nm. The first experiments that are planned 
are to determine the onset of the Coulomb explosion of particles in an 
aerosol, as a function of pulse fluence, as measured by diffraction from the 
particles by the x-ray pulses themselves.

\section{Conclusions}

We have discussed advantages and limitations of the CXDI technique, and our 
progress in implementing the experiments and reconstruction techniques. 
Preparation of special objects containing an isolated gold ball near an 
unknown compact object are shown experimentally to allow simple (although 
low resolution) image reconstruction without iteration based on the 
autocorrelation function (Fourier transform holography). Such an image may 
also provide a support function for the retrieval of higher-resolution 
images using the HIO algorithm. We have developed an algorithm that obtains 
the support automatically from the autocorrelation function by directly 
Fourier transforming the recorded diffraction intensity. This eliminates the 
need for an auxiliary imaging experiment to obtain the support function.

Our last experiment shows that the exposure times for 10 nm resolution are 
reasonable (about 10 hours) and this resolution is near the dose limit for 
life-science experiments. Thus we conclude that 3D 10-nm resolution life 
science experiments on frozen hydrated samples will be possible, and even 
better resolutions might be achievable with labeling techniques. 
Calculations indicate that a factor of about $10^4$ improvement in imaging 
speed can be expected using a dedicated CXDI beam line at the upgraded ALS. 
This is important for applications in material science where the dose 
limitation is much less severe. Based on the inverse fourth power scaling of 
the required flux with resolution, this should allow a factor ten 
improvement in resolution to about 1~nm.

These results have prompted us to develop two parallel projects.
 One is the development of a program for cryo-CXDI at ALS, and 
the other aims at femtosecond imaging of single molecules at the LCLS and 
other $4^{th}$ generation sources.

The work in CXDI at ALS will be performed in collaboration with
 groups from NSLS and Stony Brook, who have constructed an 
instrument designed for collecting 3D data sets \cite{beetz:2003}, and 
which will be installed at ALS. 
 The instrument incorporates a set of in-vacuum stepper 
motors for precision positioning the required apertures, beam stop, and 
diagnostics. It incorporates an air-lock and precision goniometer (developed 
for electron tomography) for easy introduction and manipulation of the 
specimen. The specimen holder is designed so as to be able to rotate the 
specimen over the angular range from -80 to +80 degrees before the specimen 
supporting grid obscures the beam. A unique feature of the instrument is 
that a zone plate can be positioned between the specimen and the CCD, 
thereby a low resolution image of the specimen can be recorded. This feature 
helps align the specimen, and provides information on its ``support''. 

The key idea to achieving atomic resolution imaging of radiation-sensitive 
single molecules and particles is that the damage limit may be overcome with 
the use of very short x-ray pulses to capture the data before damage occurs. 
Three problems need to be assessed to be able to perform single molecule 
imaging experiments: femtosecond damage and experiment modeling; image 
orientation, characterization and reconstruction; and single molecule sample 
handling. 

Fast and accurate hydrodynamic models to describe the interaction of a 
molecule with femtosecond x-ray pulses needs confirmation from experimental 
results of short pulse, high-field x-ray-matter interaction experiments 
planned in the near future. The recorded diffraction patterns must be 
classified, averaged to increase the signal-to-noise ratio, and oriented for 
the final 3D reconstruction. Methods developed for single-particle electron 
microscopy can be applied here. We must also develop methods to inject 
samples into the beam and, if possible, orient the particles (or at least 
influence their orientation). Preliminary lower resolution experiments on 
single particles will provide the answers to the experimental needs to build 
an experiment at LCLS to perform single-molecule imaging.

We have established the fundamental boundaries of applicability of CXDI to 
problems in materials and biological sciences. With the rapidly growing 
importance of nanostructures, and the potential that future developments in 
this area have for major breakthroughs in the fundamental and applied 
sciences and in technology, the addition of a new probe that can image thick 
objects at very high 3d spatial resolution will have a decisive impact on 
nanoscience and technology. Even more ambitiously, atomic-resolution CXDI at 
an XFEL, has the potential to provide atomic structure determination of 
biological systems. 

\begin{acknowledgments}
We would like to thank Rick Levesque of LLNL for designing and building the 
rotation stage, and Cindy Larson and Sherry Baker (both LLNL) for SEM 
imaging. This work was performed under the auspices of the U.S. Department 
of Energy by the Lawrence Livermore National Laboratory under Contract No. 
W-7405-ENG-48 and the Director, Office of Energy Research, Office of Basics 
Energy Sciences, Materials Sciences Division of the U. S. Department of 
Energy, under Contract No. DE-AC03-76SF00098. SM acknowledges funding from 
the National Science Foundation. The Center for Biophotonics, an NSF Science 
and Technology Center, is managed by the University of California, Davis, 
under Cooperative Agreement No. PHY0120999.
\end{acknowledgments}

\end{document}